\documentclass{aa}  
\usepackage[symbol]{footmisc}

\usepackage{graphicx}
\usepackage{txfonts}
\begin{document} 

  \title{Constraining transport of angular momentum in stars}
   \subtitle{Combining asteroseismic observations of core helium burning stars and white dwarfs}

   \author{J.W. den Hartogh
          \inst{1,2,3}
          \and
          P. Eggenberger\inst{4}
          \and
          R. Hirschi\inst{2,3,5}
          }

   \institute{Konkoly Observatory, Konkoly Thege Mikl\'{o}s \'{u}t 15-17, H-1121 Budapest, Hungary\\
              \email{jacqueline.den.hartogh@csfk.mta.hu}
           \and 
           Astrophysics group, Lennard-Jones Laboratories, Keele University, ST5 5BG, UK
         \and
         NuGrid Collaboration, nugrid.github.io
         \and
             Observatoire de Gen\`{e}ve, Universit\'{e} de Gen\`{e}ve, 51 Ch. des Maillettes, 1290 Sauverny, Suisse
          \and 
          Kavli Institute for the Physics and Mathematics of the Universe (WPI), University of Tokyo, 5-1-5 Kashiwanoha,
277-8583 Kashiwa, Japan
             }

   \date{Received September 27, 2018; accepted January 7, 2019}

  \abstract  
  {Transport of angular momentum has been a challenging topic within the stellar evolution community, even more since the recent asteroseismic surveys.  All published studies on rotation using asteroseismic observations show a discrepancy between the observed and calculated rotation rates, indicating there is an undetermined process of angular momentum transport active in these stars.}
  {We aim to constrain the efficiency of this process by investigating rotation rates of 2.5 M$_{\odot}$ stars.}
  {First, we investigated whether the Tayler-Spruit dynamo could be responsible for the extra transport of angular momentum for stars with an initial mass of 2.5 M$_{\odot}$. Then, by computing rotating models including a constant additional artificial viscosity, we determined the efficiency of the missing process of angular momentum transport by comparing the models to the asteroseismic observations of core helium burning stars. Parameter studies were performed to investigate the effect of the stellar evolution code used, initial mass, and evolutionary stage. We evolved our models into the white dwarf phase, and provide a comparison to white dwarf rotation rates.}
  {The Tayler-Spruit dynamo is unable to provide enough transport of angular momentum to reach the observed values of the core helium burning stars investigated in this paper. We find that a value for the additional artificial viscosity $\nu_{\rm{add}}$ around 10$^7$ cm$^2$ s$^{-1}$  provides enough transport of angular momentum. However, the rotational period of these models is too high in the white dwarf phase to match the white dwarf observations. From this comparison we infer that the efficiency of the missing process must decrease during the core helium burning phase. When excluding the $\nu_{\rm{add}}$ during core helium burning phase, we can match the rotational periods of both the core helium burning stars and white dwarfs. }
  {}


\maketitle
%

\section{Introduction}
\label{sec:introduction}
The inclusion of rotation in stellar evolution codes is complicated, due to the limited understanding of how it  affects the stellar structure \citep{2009_book_Maeder}. Some observational constraints are available from studying surface abundances and surface rotation values. These help with the calibration of the implementation of rotation in stellar evolution codes affecting both the mixing of chemical elements and the transport of angular momentum \citep{2000ApJHeger,2013meynet}. In this paper, we  use a set of core and envelope rotation rates from core helium burning stars as an extra set of observational constraints to study the transport of angular momentum.\\
A big step forward in the determination of internal rotation rates has been provided by the Kepler spacecraft \citep[see][]{Borucki2010}, as mixed modes were found in the spectra. Mixed modes are able to propagate through both convective and radiative zones, and hold information about both. The rotational frequency splittings of mixed models were first measured by \cite{2012beck} for the red giant KIC 8366239. They were then able to rule out solid-body rotation for this star. Another conclusion of this paper was that the core rotates about ten times faster than the envelope. More papers on internal rotation rates followed, also mainly focussing on observations of low-mass, evolved stars, and supported these conclusions: \citet{Deheuvels2012,Deheuvels2014,Deheuvels2015,Deheuvels2017} analyse in great detail small data sets to obtain both core and surface rotation rates, while \citet{2012Mosser,Gehan2018} analyse the core rotation rates of a data set of several hundred giants (and core helium burning) stars.\\
Stellar evolution codes have been unable to match the observed ratio of core and envelope rotation rates. Predicted core rotation rates are about two orders of magnitude higher than observed rates \citep[see][]{2012eggenberger,Marques2012}. This is a major issue within the stellar evolution community as it shows that a process for transport of angular momentum is missing from the current implementation of rotation. \\
Broadly speaking, there are three types of processes which could improve the transport of angular momentum in stars: hydrodynamical, wave-driven, and magnetic. There is no consensus on which process dominates. \citet{Eggenberger2005,2008Suijs,2014cantiello} show that magnetic fields, and the Tayler-Spruit (TS) dynamo in particular \citep{Spruit1999,Spruit2002} are effective in coupling the core and envelope to increase the transport of angular momentum. While the TS dynamo is able to reproduce the solar rotation profile, it cannot provide the coupling needed to match the asteroseismic observations of red giants. Also, the inclusion of the TS dynamo has not been tested for the full mass range for which we have observational constraints on the internal stellar rotation rates. \\
The transport of angular momentum by low-frequency internal gravity waves has been studied with multi-D simulations \citep{2014fuller,Rogers2015,Rogers2017}. They conclude that internal gravity waves are a promising method for transporting angular momentum. However its behaviour is complex and difficult to translate to a 1D parametrisation that can be included in 1D stellar evolution codes. \citet{Pincon2017} showed that, for red giants, the gravity waves alone do not transport enough angular momentum to match observations. However, they could provide the needed transport during the subgiant phase. \citet{2015belkacem} show that the transport by mixed models could play a role for evolved red giants.\\
For  the magnetic fields and for the wave-driven option, there is still work to be done on the physical process itself. Therefore, it is important to derive the efficiency of the missing process of angular momentum so that its physical character can be revealed. To this end, \cite{2012eggenberger} and \citet{Eggenberger_2017} have performed stellar evolution calculations including both hydrodynamical processes like shear and meridional circulation, and a constant additional artificial viscosity ($\nu_{\rm{add}}$), that only influences the angular momentum transport and not the chemical composition mixing. This constant has no physical meaning; it is added to investigate the level of efficiency the missing process of angular momentum transport should have, and is used to investigate whether we can determine the evolution of the missing process. \citet{2012eggenberger} focus on the 1.5 M$_{\odot}$ red giant KIC8366239 observed by \citet{2012beck} and the efficiency of the missing process needed to match those observations. The authors found a $\nu_{\rm{add}}$ of 3$\times$10$^4$ cm$^2$ s$^{-1}$ as a mean value for the efficiency of the transport process, constrained strongly by the asteroseismic observations. \citet{2016spada} followed a similar approach to constrain the missing process by including a diffusive process to the transport of angular momentum that varies with the ratio of core to envelope rotation rate, inspired by the azimuthal magneto-rotational instability \citep[AMRI, see][]{2007rudiger}. They compare their stellar evolution models to observations of core rotation rate from \citet{Deheuvels2014} and \citet{2012Mosser} and conclude that the missing process of angular momentum transport has to change throughout the evolution of a star to be able to match the post-main sequence rotational evolution of low-mass stars. \cite{Eggenberger_2017}, following the same strategy as \cite{2012eggenberger}, focussed on KIC7341231, a 0.84 M$_{\odot}$ red giant, for which \citet{Deheuvels2012} deduced a core rotation rate and an upper limit for the surface rotation rate. For this red giant  an additional transport process was again needed, and the authors determined the efficiency to be 1$\times$10$^3$ < $\nu_{\rm{add}}$ cm$^2$s$^{-1}$ < 1.4$\times$10$^4$. This value is lower than that found for the efficiency of the missing transport process for the more massive KIC8366239, so \citet{Eggenberger_2017} concluded that the missing process of angular momentum transport is sensitive to both evolutionary phase and the initial mass. \\
This paper focusses on further constraining the missing process of angular momentum transport by adding a $\nu_{\rm{add}}$ to stellar evolution calculations. The asteroseismic observations used are the seven core helium burning stars from \citet{Deheuvels2015}. These stars have a mass around 2.5 M$_{\odot}$, which means they have not experienced helium flashes during their evolution, and therefore the evolution of their rotational properties post-main sequence are different from KIC7341231 and KIC8366239. For each of the seven stars,  the surface and core rotation rates are both published. This allows us to put strong constraints on the efficiency of the missing process of angular momentum transport. The structure of the paper is as follows. In Sect. 2 we introduce our methodology, and in Sect. 3 we describe the evolution of our 2.5 M$_{\odot}$ models up to the core helium burning phase. In Sect. \ref{sec:ts-dynamo} we investigate whether the TS dynamo is also unable in this case  to provide the coupling needed to match observations. In Sect. \ref{sec:artvisc} we determine the efficiency of the missing process of angular momentum transport needed to match the rotation rates obtained by \citet{Deheuvels2015}. Section \ref{sec:WDs} follows with the comparison of our models to the observed white dwarf rotation rates. We end the paper with our conclusions in Sect. 6.


\section{Physics of the models}
The stellar evolution calculations presented in this paper were performed with the Module for Experiments in Stellar Astrophysics (MESA). The MESA code is described in the code papers \citep[see][]{2011_MESA_1,2013_MESA_2} and will not be repeated here. Most initial parameters match the papers of the Nugrid collaboration  \citep{2016Pignatari,Battino_2016}. In summary, the Schwarzschild criterion is used for the convective boundary placement, while exponentially decaying convective boundary mixing as introduced by \citet{1997herwig_overs} is used for the boundaries. Mass loss on the red-giant branch (RGB) is set according to \citet{1975reimers}, on the AGB by \citep{1995blocker}. OPAL Type 2 opacities are used \citep{1996rogers}, and for the lower temperatures \citet{Ferguson2005}.
\subsection{Rotation}
The implementation of rotation in MESA follows \cite{2000ApJHeger} and we use the default settings as defined in that paper. The transport processes included in calculations in this paper are the Eddington-Sweet circulation, dynamical and secular shear instabilities, the Solberg-H{\o}iland criterion  and the TS dynamo. We excluded the Goldreich-Schubert-Fricke instability because both \cite{2010Hirschi} and \cite{2016caleo} show that the GSF instability is not likely to  contribute to the transport of angular momentum and might not be present at all in stars. Rotation is included at the zero age main sequence (ZAMS) as rigid body rotation. Although the implementation of rotation in MESA is different from that in GENEC, the code used by Eggenberger et al. (2012,2017), we confirm that we find similar $\nu_{\rm{add}}$ to explain the observations as in those papers (see Appendix \ref{sec:model_unc} for a comparison between models of the two codes).\\
Several techniques exist in MESA to smooth the diffusion profiles of the instabilities, individually or their sum. We did not use any of these techniques, apart from a  technique that smoothes the diffusion profile of the TS dynamo over time, exactly as  was included by \cite{2014cantiello}. Without this smoothing technique, the stellar evolution calculation with and without the TS dynamo show rotation rates in the same order of magnitude.\\
In MESA, to calculate the transport of angular momentum, we use 
\begin{equation}
\left( \frac{\partial \Omega}{\partial t}\right)_m=\frac{1}{j}\left( \frac{\partial}{\partial m}\right)_t \left[ (4\pi r^2 \rho)^2jD_{\rm{am}} \left( \frac{\partial \Omega}{\partial m}\right) \right]-\frac{2\Omega}{r}\left(\frac{\partial r}{\partial t}\right)_m\left( \frac{1}{2}\frac{d\rm{ln}j}{d\rm{ln}r}\right)   
\label{eq:AMtranport}
,\end{equation}
where $\Omega$ is the angular velocity, $j$ the specific angular momentum, and total diffusion coefficient $D_{\rm{am}}$ takes into account all processes that transport angular momentum, including the hydrodynamical ones listed previously. The $\nu_{\rm{add}}$ is also added to this variable D$_{\rm{am}}$. This implementation is identical to the implementation of the $\nu_{\rm{add}}$ by Eggenberger et al. (2012, 2017). At the inner and outer boundary $\partial \Omega/\partial t$ is set to 0, following \citet{2000ApJHeger}.\\
The general evolution of the structure and angular momentum profiles of the models are presented in  Appendix \ref{sec:evo_rot} Uncertainties on the models are discussed in Appendix \ref{sec:model_unc}.

\subsection{The seven KIC stars}
\label{sec:7KIC}
In Table \ref{tab:deheuvels} we summarise the important parameters of the seven KIC stars used as comparison sample, which are taken from \citet{Deheuvels2015}. We include the core and surface rotation rates, the ratio between them, and the surface gravity (log g), all with their error margins. The metallicities of the seven stars are around solar, according to the APOGEE Data Release 14 \citep{Abolfathi2018}, which includes all seven stars. We used a metallicity of Z = 0.014 and the metal abundance mixture of \cite{1993grevesse}, and therefore focus on matching the global trends of the seven stars as a group instead of trying to find best-fit models for each star individually. This allows us to constrain the missing process of angular momentum for core helium burning stars.\\
The initial mass of our models is chosen to be 2.5 M$_{\odot}$ because this is very close to the mean observed mass of the seven KICs. In Appendix \ref{sec:model_unc} we will see that the influence of the stellar evolution code used on the rotational properties is negligible.\\
Other observations of rotation rates in evolved stars in the same mass range have been published in \citet{Massarotti2007}, \citet{2012Mosser}, \citet{Tayar2015}, and \citet{Ceillier2017} and analysed in \citet{Tayar2018}. These data sets, however, only include either the surface or the core rotation rates. To date, the data set of \citet{Deheuvels2015} is the only data set in the 2 to 3 M$_{\odot}$ mass range that provides both rotation rates. This allows us to constrain our models better than when we only have one of the rates, so we only use the data set of \citet{Deheuvels2015} in this study.  \\
It is important to note that the rotation rates labelled as `core' rotation rates are actually `near core' values, as shown in Fig. 5 of \citet{Deheuvels2015}. The comparison of the calculations to the region where the observations of the core rotation originate from is explained in Appendix C. 
   
\begin{table*}
\centering
\caption{Properties of the seven KIC stars from \citet{Deheuvels2015}. From left to right we list the Kepler Input Catalog ID, the obtained mass, surface gravity and rotation rates of core and envelope. The last column shows the ratio of the rotation rates.}
\begin{tabular}{c | cccccc}
\hline 
KIC-id & M/M$_{\odot}$ &  log$_{10}$($g$/cm s$^{-2}$) & $\Omega_{\rm{c}}$/(2$\pi$ nHz) & $\Omega_{\rm{s}}$/(2$\pi$ nHz)   & $\Omega_{c}$/$\Omega_{s}$  \\ 
\hline 
KIC5184199 & 2.18 $\pm$ 0.23 & 2.907 $\pm$ 0.012 & 200 $\pm$ 13 & 63 $\pm$ 20 & 3.2 $\pm$ 1.0  \\ 
KIC4659821 & 2.21 $\pm$ 0.18 & 2.935 $\pm$ 0.013 & 165 $\pm$ 14 & 79 $\pm$ 15 & 2.1 $\pm$ 0.4  \\ 
KIC8962923 & 2.23 $\pm$ 0.26 & 2.832 $\pm$ 0.013 & 138 $\pm$ 8 & 79 $\pm$ 10 & 1.8 $\pm$ 0.3  \\ 
KIC3744681 & 2.45 $\pm$ 0.35 & 2.712 $\pm$ 0.015 & 194 $\pm$ 20 & 63 $\pm$ 36 & 3.1 $\pm$ 1.8  \\ 
KIC9346602 & 2.51 $\pm$ 0.36 & 2.675 $\pm$ 0.013 & 164 $\pm$ 6 & 53 $\pm$ 15 & 3.1 $\pm$ 0.9  \\ 
KIC7467630 & 2.57 $\pm$ 0.27 & 2.776 $\pm$ 0.015 & 121 $\pm$ 18 & 96 $\pm$ 28 & 1.3 $\pm$ 0.4  \\ 
KIC7581399 & 2.90 $\pm$ 0.34 & 2.843 $\pm$ 0.013 & 164 $\pm$ 12 & 87 $\pm$ 14 & 1.9 $\pm$ 0.3  \\ 
\hline 
\end{tabular} 
\label{tab:deheuvels}
\end{table*}

\section{Can the TS dynamo provide enough coupling to explain asteroseismic derived rotation properties of core helium burning stars?}
\label{sec:ts-dynamo}
\begin{figure}
   \centering
   \includegraphics[width=\linewidth]{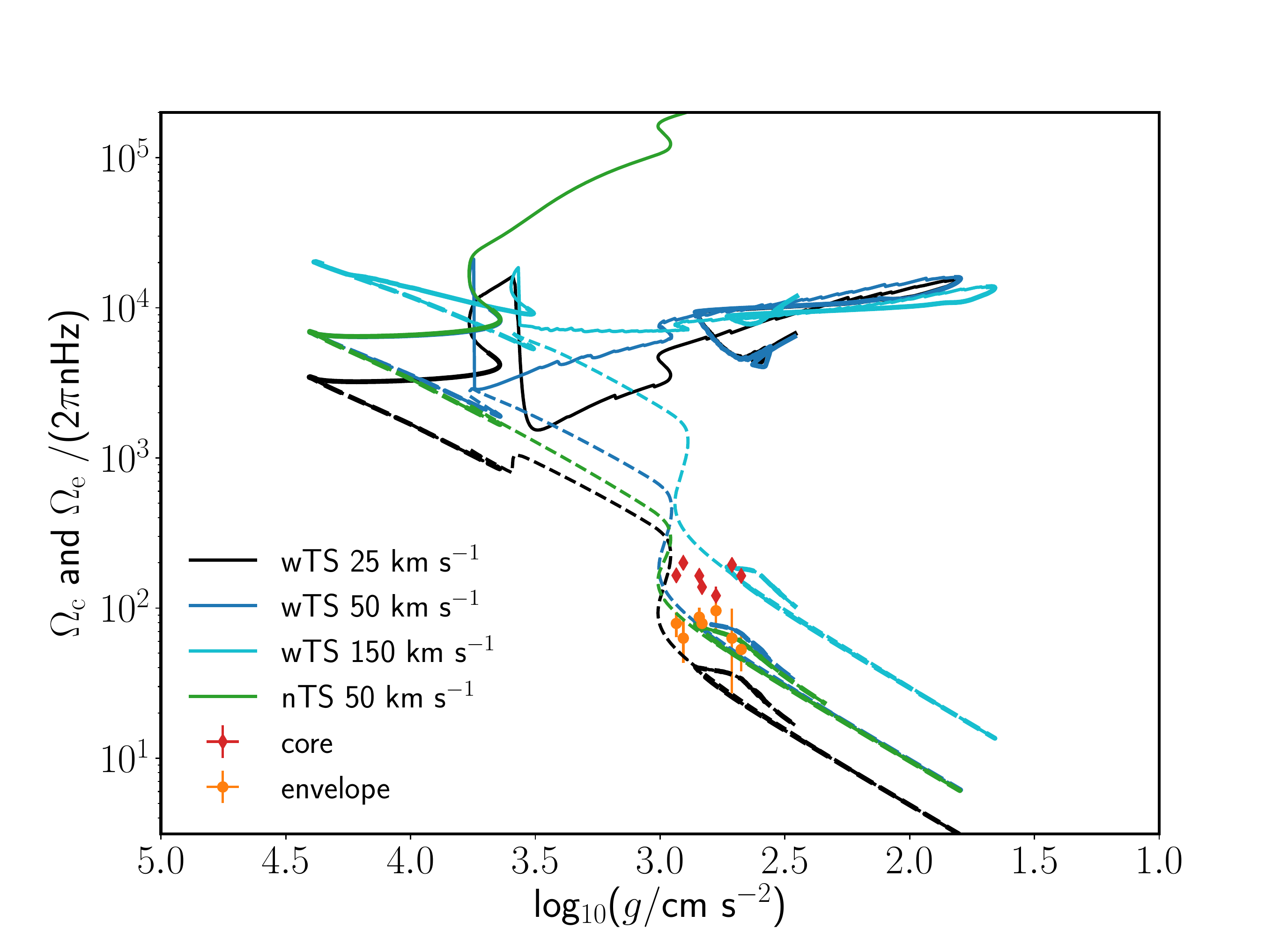} 
   \caption{Effect of inclusion TS dynamo. Rotation rates of the core (solid line) and envelope (dashed line) of the models with the TS dynamo (wTS) and without (nTS). The initial rotation rates of the models are included in the legend. Data points are from \cite{Deheuvels2015}. }
   \label{fig:omegaR-BR+B}%
   \end{figure}

The first goal of this paper is to investigate whether the TS dynamo provides enough coupling between core and envelope to match the observations of the core helium burning stars analysed by \cite{Deheuvels2015}. \cite{2014cantiello} show, for stars with an initial mass of 1.5 M$_{\odot}$, that during the early RGB  inclusion of the TS dynamo provides more coupling between core and envelope but not enough to match the RGB rotation rates provided by \cite{2012Mosser}. Thus, they concluded that the RGB phase is the evolutionary phase where more coupling is needed. However, the evolution of 1.5 M$_{\odot}$ and 2.5 M$_{\odot}$ are very different, in particular during the RGB phase. Stars with an initial mass below about 2 M$_{\odot}$ undergo helium flashes in the core after it has become degenerate, and cores of stars with a higher initial mass ignite core helium burning before becoming degenerate. As a consequence, the times between the end of core hydrogen and the start of core helium burning are different; our calculations show a difference of one order of magnitude. For this reason, testing the conclusions of \cite{2014cantiello} for 2.5 M$_{\odot}$ stars is a valuable task, especially when comparing them with observations of stars that are already past the RGB phase.\\
Figure \ref{fig:omegaR-BR+B} shows the core (solid line) and envelope (dashed line) rotation rates of our 2.5 M$_{\odot}$ models as a function of the surface gravity with different initial rotational velocities: 25, 50, and 150 km/s. The start of the main sequence (MS) is where the core and envelope rotation rates are equal (top left)  and the end of the core helium burning phase is where core and surface rotation rates are the furthest apart (middle and bottom right). The core H and core He burning phases are both shown in thick line widths, while the RGB phase is shown in thinner line width. Starting with the comparison of the surface rotation rates (dashed lines), we see that the 50 km/s models, with the TS dynamo (wTS) and without (nTS), reach five of the seven data points, while the 25 km/s model reaches one of the seven and the 150 km/s model  reaches none. We therefore set the initial rotation rate of all the models to 50 km s$^{-1}$. The two other data points can be reached by reducing the initial mass of the models, see Appendix \ref{sec:model_unc}.\\
When focussing on core rotation rates during the core helium burning phase, we see that all models including the TS dynamo ($\Omega_{c}\simeq$10$^{4}$ nHz) are two orders of magnitude away from the data points. Including the TS dynamo  improves the match to the observations as the difference between observations and the model without the TS dynamo ($\Omega_{c}\simeq$10$^{6-7}$ nHz) is more than 3 orders of magnitude worse. We thus conclude that also for the 2.5 M$_{\odot}$ stars, the TS dynamo does not provide enough coupling between core and envelope to reduce the core rotation rates enough to match asteroseismic observations of the core helium burning stars. \\

\section{Additional viscosity needed to reproduce observations of helium burning stars}
\label{sec:artvisc}
Now that we have shown that the models with and without the TS dynamo cannot reproduce the asteroseismic observations of the seven secondary clump stars from \cite{Deheuvels2015}, we continue by determining the strength of the missing process of angular momentum transport as a first step to revealing its physical nature. To do so, a constant $\nu_{\rm{add}}$ is added to the transport of angular momentum. We  stress, however,  that we do not believe the missing process of angular momentum transport is constant.

\subsection{Determination of the additional viscosity needed to reproduce the \cite{Deheuvels2015} data}
\label{sec:determine}

\begin{figure}
   \centering
   \includegraphics[width=\linewidth]{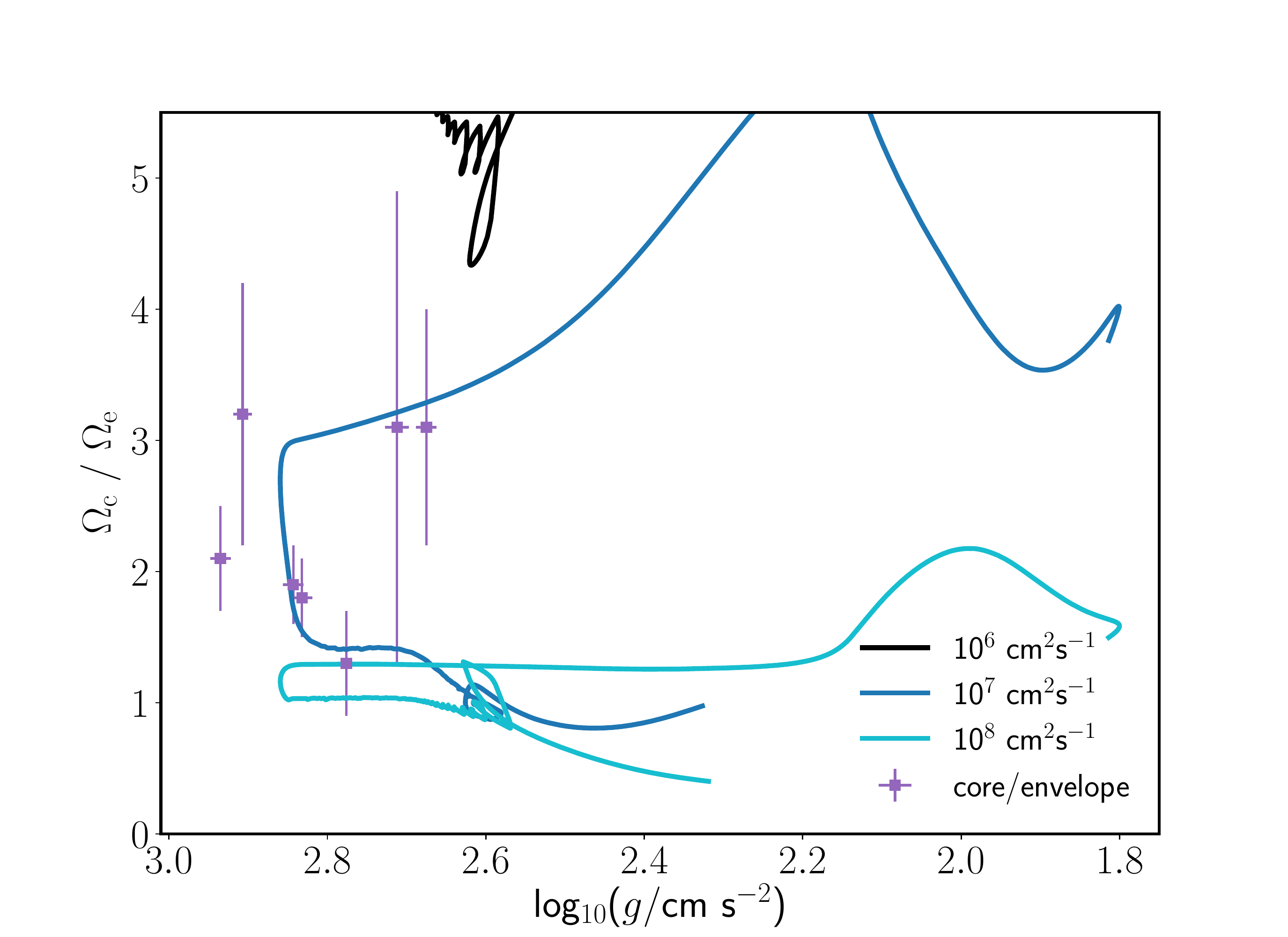}
   \caption{Determining the $\nu_{\rm{add}}$. The ratio of core to surface rotation rate as a function of surface gravity for three models calculated with an initial rotational velocity of 50 km s$^{-1}$, while the $\nu_{\rm{add}}$ is varied. The data points are from \citet{Deheuvels2015}. }
   \label{fig:ratio_artvisc}%
   \end{figure}

\begin{figure}
   \centering
   \includegraphics[width=\linewidth]{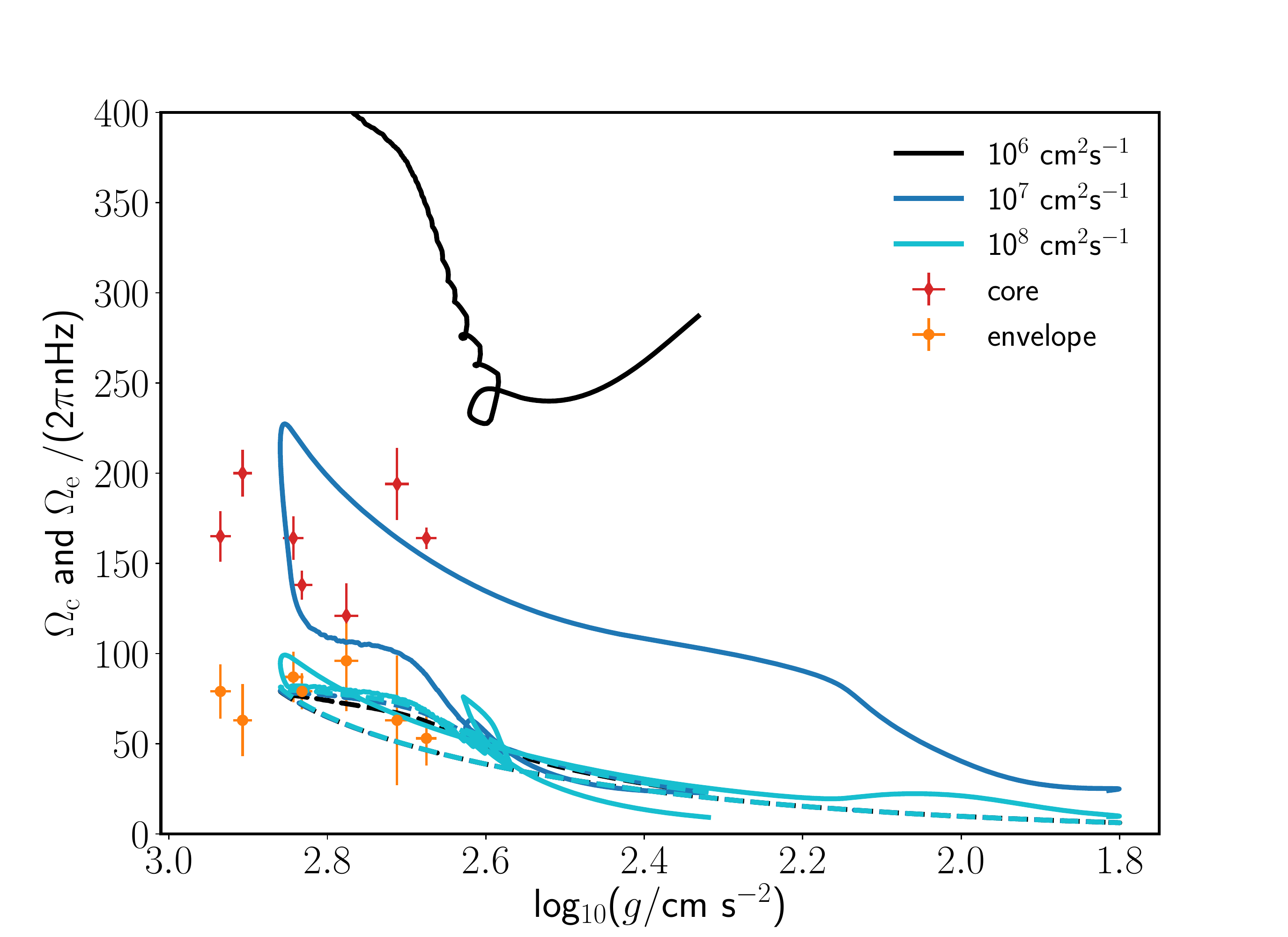}
   \caption{Determining the $\nu_{\rm{add}}$. Core and surface (solid and dashed line, respectively) rotation rates as a function of surface gravity for three models in Fig. \ref{fig:ratio_artvisc}. The data points are from \citet{Deheuvels2015}.}
   \label{fig:omega_artvisc}%
   \end{figure}

From \cite{Eggenberger_2017} we know that the efficiency of the unknown transport process for angular momentum increases with stellar mass. Therefore, in this study a stronger process is expected than employed by \cite{2012eggenberger}, studying a 1.5 M$_{\odot}$ star, and \cite{Eggenberger_2017}, studying a 0.84 M$_{\odot}$ star. \\
As mentioned before, we did not attempt to fit all stars separately, but we look for global trends instead. Using Fig. \ref{fig:ratio_artvisc}, we determined the global efficiency of the missing process of angular momentum in the seven KIC stars. Figure \ref{fig:ratio_artvisc} shows the ratio of core to envelope rotation rate, which, as mentioned by \cite{Eggenberger_2017}, allows us to determine $\nu_{\rm{add}}$ independently of the initial rotation rate. The best match in Fig. \ref{fig:ratio_artvisc} is $\nu_{\rm{add}}$=10$^7$ cm$^2$ s$^{-1}$, which matches five of the seven data points. The other two models included reach none ($\nu_{\rm{add}}$=10$^6$ cm$^2$ s$^{-1}$) or two ($\nu_{\rm{add}}$=10$^8$ cm$^2$ s$^{-1}$) of the data points. More importantly, the general trend shown by the data points is best matched by the model that includes a $\nu_{\rm{add}}$ of 10$^7$ cm$^2$ s$^{-1}$. Again, the two data points with the highest surface gravities cannot be reached (see Appendix \ref{sec:model_unc} for how to reach these points).\\ 
When comparing the lines in Fig. \ref{fig:omega_artvisc} to the lines in Figs. \ref{fig:log_g_15_101} and \ref{fig:log_omega_15_101}, we can determine the start of the core He burning phase in Fig. \ref{fig:omega_artvisc}. This is at the lowest surface gravity, in the bottom right corner of the figure. Then, both surface gravity $g$ and the core rotation rate $\Omega_{\rm{c}}$ increase in a short amount of time until steady core He burning sets in and a slow decrease in both surface gravity and the core rotation rate characterises the rest of this phase. All data points are positioned around the turning point of the trend in surface gravity. From Fig. \ref{fig:log_omega_15_101} it follows that these seven stars are thus in the early phases of core He burning.\\
In Fig. \ref{fig:omega_artvisc} the core and surface rotation rates are shown for the same three models as in Fig. \ref{fig:ratio_artvisc}. This figure confirms the choice for the initial rotation rate because the data points for surface rotation are matched. Also in this comparison, the general trend shown by the data points is best matched by the model with a $\nu_{\rm{add}}$ of 10$^7$ cm$^2$ s$^{-1}$.

\subsection{Time dependence of the additional viscosity}
\begin{figure}
   \centering
   \includegraphics[width=\linewidth]{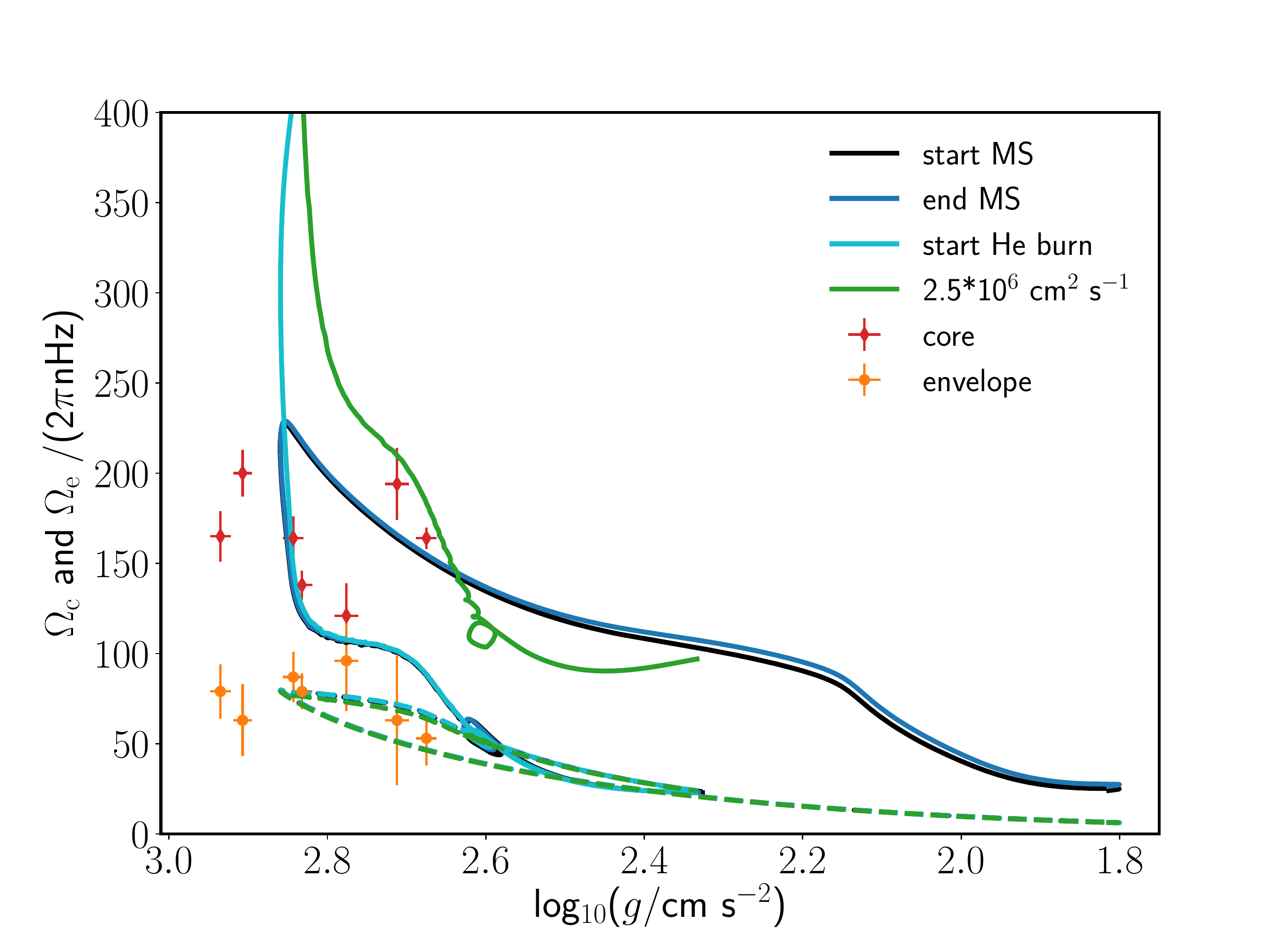}
   \caption{Dependence on inclusion time. The models presented here have been calculated with the best fit parameters (50 km s$^{-1}$, 10$^7$ cm$^2$ s$^{-1}$), apart from the model labelled  `2.5*10$^6$ cm$^2$ s$^{-1}$', while varying the moment of including the $\nu_{\rm{add}}$. The labels reflect the phase when the $\nu_{\rm{add}}$ is included. For the 2.5 10$^6$ cm$^2$ s$^{-1}$ model, the moment of inclusion is at the start of the core helium burning phase.}
   \label{fig:omega_artvisc_3moments}%
   \end{figure}

In Sect. \ref{sec:determine} we showed that the mean efficiency of the missing transport mechanism in the seven stars of \citet{Deheuvels2015} is around 10$^7$ cm$^2$ s$^{-1}$ when adding the $\nu_{\rm{add}}$ at the start of the main sequence. In this section we investigate whether this is dependent on the evolutionary phase during which $\nu_{\rm{add}}$ is added to the calculation. By doing this we are able to determine whether there is a phase in which the transport of angular momentum dominates the rest of the evolution. In this section we focus on the evolution up to the core helium burning phase and in Sect. \ref{sec:wd_evo} we focus on the later phases to investigate the influence of the inclusion of $\nu_{\rm{add}}$ on the final white dwarf spin.\\
We calculated models that include the $\nu_{\rm{add}}$ only from the end of the main sequence and from the start of the core helium burning phase. For the first, we find that adding the same $\nu_{\rm{add}}$ is sufficient to reach the data points, see Fig. \ref{fig:omega_artvisc_3moments}, and that this model is comparable to the model in which we included $\nu_{\rm{add}}$ from the start of the main sequence. Therefore, we conclude that the main sequence is not a dominant phase for angular momentum transport in our models and we have no arguments to exclude $\nu_{\rm{add}}$ during the main sequence either. \\
The inclusion of $\nu_{\rm{add}}$ only at the start of the core helium burning phase changes the evolution of the rotation rates, see again Fig. \ref{fig:omega_artvisc_3moments}. Without the $\nu_{\rm{add}}$ earlier in the calculation, the core rotation rate is higher at the start of the core helium burning phase in this model than in the models that do include $\nu_{\rm{add}}$ earlier in the evolution. This is why the line of this model starts at a different point (top-left corner) in the figure. However, this difference has disappeared around log$_{10}$($g/$cm s$^{-2}$)$\simeq$ 2.8. The location of the curve in this model is dependent on $\nu_{\rm{add}}$, shown by the model labelled `2.5 10$^6$ cm$^2$ s$^{-1}$', this number being the $\nu_{\rm{add}}$ added at the start of the core helium burning phase. Thus, when we add the $\nu_{\rm{add}}$ at the start of the core helium burning phase, we are still able to reach all data points. However, we then have to use a value in the range of 2.5 10$^6$ < $\nu_{\rm{add}}$ cm$^2$ s$^{-1}$ < 10$^7$ . While the data cannot rule out the models that include $\nu_{\rm{add}}$ at the start of the core helium burning phase, the data does favour earlier inclusion of $\nu_{\rm{add}}$ because no data points are found with an angular velocity of the core above 200 nHz. \\
When we suppress the $\nu_{\rm{add}}$ from the start of the core helium burning phase onwards, we are unable to reach any data points. The reason for this is that the molecular weight gradient is too strong and without any $\nu_{\rm{add}}$ there is no transport of angular moment over this gradient. Therefore, we conclude that the crucial phase for the transport of angular momentum is the start of the core helium burning phase. \\

\section{White dwarf rotation rates}  
\label{sec:WDs}
After the core helium burning phase, we continued the models until they reached the white dwarf phase. In between these two phases, the stars pass through the asymptotic giant branch phase (AGB). During this phase, the energy production comes from the hydrogen and helium burning shell, located between the core and envelope. The helium shell becomes unstable, resulting in thermal pulses (TP-AGB phase). Around 25 to 30 thermal pulses take place in this phase in our models, and  between each TP a third dredge-up (TDU) can occur. During the TP-AGB phase mass loss is enhanced, leading to removal of the envelope. Via the planetary nebulae phase, the star moves to the white dwarf track.
\subsection{Calculation of the AGB phase}
We calculated the full AGB phase as we would have done when studying the s-process nucleosynthesis \citep[see][for details]{2016Pignatari,Battino_2016}. For instance, for the mass-loss treatment during the AGB phase we used \citet{1995blocker} with an efficiency of 0.01 at the start of the AGB phase, 0.04 from when the envelope is carbon rich, and to 0.5 when the convergence issues appear (see below). We also used calibrated parameters for convective boundary mixing specifically for the AGB phase.\\
This is an improvement compared to the works of \citet[][no AGB specific mass loss, manually stopped models somewhere in AGB phase]{2008Suijs}, \citet[][no details given apart from initial mass and rotational velocity]{2013tayar}, and \citet[][unphysical large mass loss efficiencies in the AGB phase which shorten this phase]{2014cantiello}. By calculating the whole AGB phase, we can investigate the effects of the $\nu_{\rm{add}}$ on the thermal pulse cycle by investigating both the transport of angular momentum and the s-process nucleosynthesis, and compare them to the standards models without $\nu_{\rm{add}}$. Details will be presented in a forthcoming paper. We report that the models with $\nu_{\rm{add}}$ included during the TP-AGB phase are able to transport angular momentum during the TDUs. This is due to the TDU reducing the molecular weight gradient and therefore the (local) barrier that has to be overcome to transport angular momentum.\\
It is common for convergence issues to arise during the final TPs in calculations like these and we report that these issues also occur in all models presented in this paper. There are two options for how to proceed, the first being the continuation of the AGB phase with a higher mass loss rate and the second the ejection of the whole remaining envelope \citep[see][]{Wood1986,Herwig1999,Sweigart1999,Lau2012}. We proceed with the models by increasing the mass loss parameter from 0.04 to 0.5, which allows for a smooth continuation of the models into the white dwarf phase.

\subsection{Final spins of best fit models}
\label{sec:wd_evo}

\begin{figure*}
   \centering
   \includegraphics[width=\linewidth]{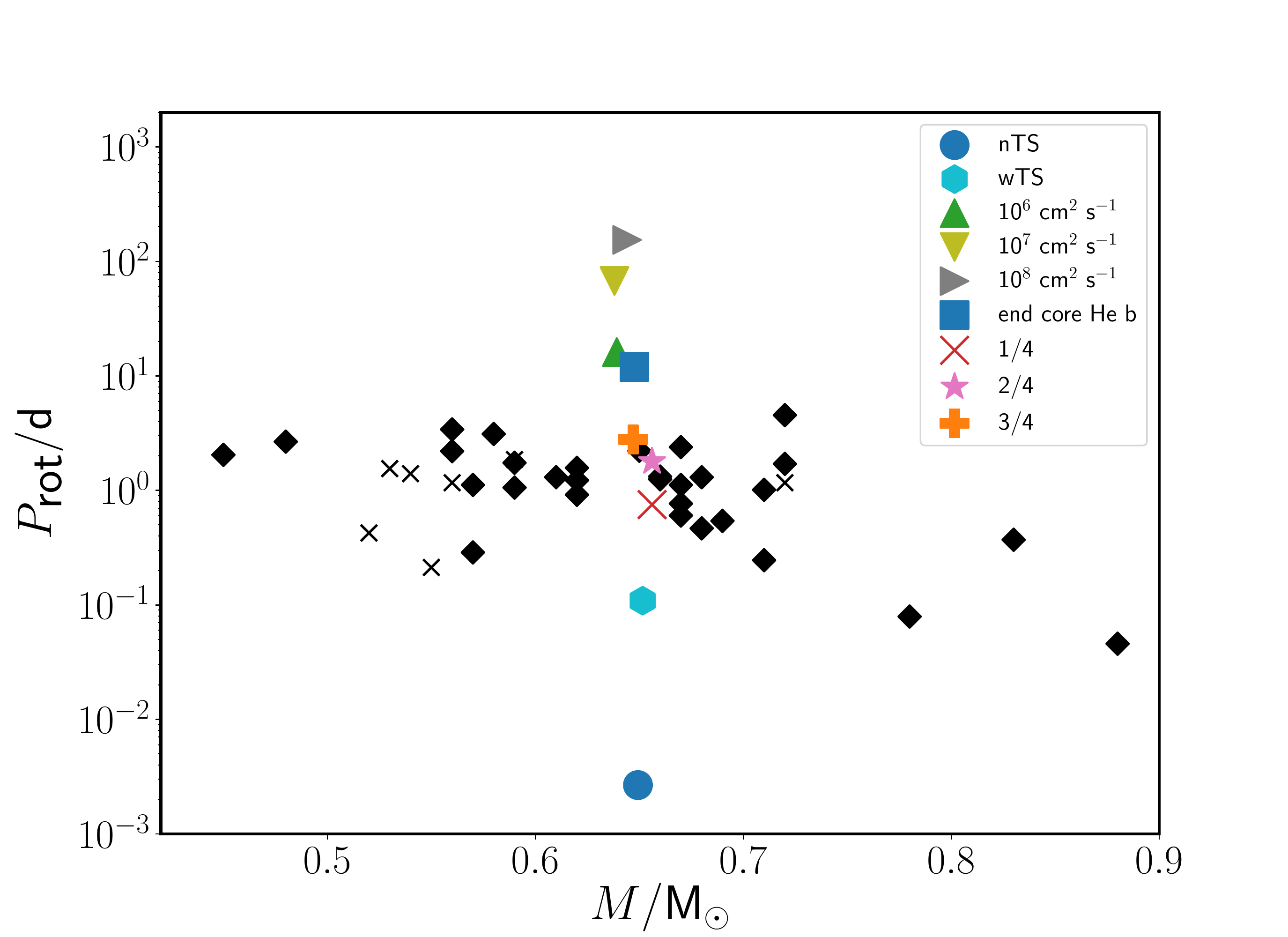}
   \caption{WD periods as a function of mass. The data points are from \cite{2015kawaler} and \cite{Hermes2017}. The black diamonds are the DAVs, the black crosses are other pulsating white dwarfs. All coloured symbols are our predicted WD periods: the sphere and hexagon are the models without $\nu_{\rm{add}}$; the triangles are the models with different values of $\nu_{\rm{add}}$; the square is the model that excludes $\nu_{\rm{add}}$ at the end of core helium burning; and the star, cross, and plus signs are the models that exclude $\nu_{\rm{add}}$ at different times during the core helium burning phase.}
   \label{fig:WDspins}%
   \end{figure*}

In this section we show the comparison between the calculated rates and the observed white dwarf rotation rates by \citet{Hermes2017} and the compilation by \citet{2015kawaler}. Most pulsating white dwarfs (WDs) in these two papers are DAVs, variable WDs with spectral type DA having only hydrogen absorption lines in their spectra. These pulsating WDs can be found in a specific temperature regime where their surface hydrogen has to become partially ionised. This regime for about 0.6 M$_{\odot}$ white dwarfs is between 12 600 and 10 600 K, and we show the rotational periods of our models when passing through that same temperature regime in Fig. \ref{fig:WDspins}. The observational points from other pulsating white dwarfs are depicted as black crosses, while the DAVs are shown as black diamonds. The number of observed white dwarf periods is still low \citep[36, we removed EPIC 201730811 because it is in a post-common envelope close binary according to][]{Hermes2015}, no statistical comparison is provided. White dwarf spins are also available for magnetic white dwarfs \citep[see][for a summary]{2015kawaler}. All of our models are non-magnetic, with only one exception, so we do not include these data points in our comparison.  \\ 
All coloured symbols in Fig. \ref{fig:WDspins} are WDs from our models. The two blue symbols correspond to the models introduced in Sect. \ref{sec:ts-dynamo}, where we tested the impact of the TS dynamo. These models are the only ones without $\nu_{\rm{add}}$ in Fig. \ref{fig:WDspins}. As already shown by \citet{2008Suijs} and \cite{2014cantiello}, the model without the TS dynamo (nTS, dark blue circle) is orders of magnitude lower than the observed white dwarf periods. The model that does include the TS dynamo (wTS, light blue hexagon) reaches the lower limit of observed white dwarf periods, but as we saw before this model does not reach the observed periods of core He burning stars.\\
All models that include $\nu_{\rm{add}}$ in Fig. \ref{fig:WDspins}, have a spin period that is larger than all observed white dwarf rotation rates. There are three models with $\nu_{\rm{add}}$ of 10$^6$, 10$^7$, and 10$^8$ cm$^2$ s$^{-1}$ included during the whole calculation (three triangles), and one model where we excluded the $\nu_{\rm{add}}$ of 10$^7$ cm$^2$ s$^{-1}$ from the end of the core He burning phase (square). All these models are introduced in Sect. \ref{sec:determine}, except for the last one. From the previous section, we know that only the models labelled  `10$^7$ cm$^2$ s$^{-1}$' and `end core He b' match the core He burning observations. However, they all transport too much angular momentum in the later phases of the evolution to match the white dwarf observations. Even the model that does not include $\nu_{\rm{add}}$ after the core He burning phase is finished does not reach the observed white dwarfs periods. Therefore,  the efficiency of the missing process of angular momentum is negligible after the end of the core He burning phase according to our models, and the efficiency of the missing process also has to change during the core helium burning phase itself.\\
To investigate this last conclusion in more detail, we calculated models where we include $\nu_{\rm{add}}$ at the ZAMS and exclude it at different moments during the core helium burning phase. The whole core helium burning phase lasts for 183 Myr in these models and $\nu_{\rm{add}}$ has been excluded from times that correspond to 1/4, 2/4, and 3/4 of that time span. After excluding $\nu_{\rm{add}}$ we continue the calculation into the white dwarf phase. These three new models have also been included in Fig. \ref{fig:WDspins}. Again the rotational period within the DAV temperature range is used\footnote{Apart from model `1/4' because this model undergoes a very late thermal pulse (VLTP) during the WD phase and is rebrightened before the DAV temperature range is reached. Convergence issues prevent the model from returning to the WD phase. We therefore calculated the rotational period of this WD just before the VLTP.}. All three models are located within the range of observed white dwarf periods, and all three therefore match both the core helium burning and white dwarf observed rotation rates.

\section{Conclusions}

In this paper, we investigated the efficiency of the missing process of angular momentum by calculating 1D stellar evolution models with an initial mass of 2.5 M$_{\odot}$. As observational tests, we used the observed core and surface rotation rates of core helium burning stars as published by \citet{Deheuvels2015} and white dwarf rotational periods published by \citet{2015kawaler} and \citet{Hermes2017}. The main conclusions of this paper are the following:
\begin{itemize}
\item As for the 1.5 M$_{\odot}$ of \cite{2014cantiello}, the 2.5 M$_{\odot}$ models including the TS dynamo do not provide enough coupling between core and envelope to match asteroseismic observations of core rotation rates.
\item We have added a constant additional viscosity to our model as a first step towards revealing the physical nature of the missing process of angular momentum transport.
\item We are able to match the core rotation rates published by \citet{Deheuvels2015} by adding $\nu_{\rm{add}}$ = 10$^7$ cm$^2$s$^{-1}$ and using an initial rotational velocity of 50 km s$^{-1}$. This order of magnitude for $\nu_{\rm{add}}$ is independent of stellar evolution code, and initial mass (see Appendix \ref{sec:model_unc}).
\item The trends identified by \citet{Eggenberger_2017} concerning the increase in $\nu_{\rm{add}}$ with both initial mass and evolutionary phase are confirmed here. See Table \ref{tab:summary} for an overview of all published studies on $\nu_{\rm{add}}$. The strong increase in $\nu_{\rm{add}}$ from the two lower mass studies to this 2.5 M$_{\odot}$ study suggests that when increasing the initial mass of the star, the change from radiative to convective core has less effect on the efficiency of the missing process of angular momentum than the absence of helium flashes in the more massive stars. 
\item We show that the dynamical instabilities (DSI and SH) are not attributed to the transport of angular momentum from ZAMS to the end of core helium burning in our models (see Appendix \ref{sec:model_unc}).
\item We show that the extra transport of angular momentum that fits the observations of the core helium burning phase leads to rotation periods in the WD phase that are too high. Our results show that the efficiency of the missing process needs to change during the core helium burning phase, and must be strongly decreased before the end of the core helium burning phase. 
\item When excluding $\nu_{\rm{add}}$ at 1/4, 2/4, or 3/4 of the whole duration of the core helium burning phase, our models match the observed rotation rates of both the set of core helium burning stars and the set of white dwarfs.
\item This implies that transport processes for which the efficiency only depends on the amount of differential rotation (such as the diffusive mixing introduced in \citealt{2016spada}, based on the AMRI by \citealt{2007rudiger}) are incompatible with the result that the missing process has to be strongly decreased by the end of the core helium burning phase, unless an inhibiting effect is included to facilitate the decrease.
A consequence of this work is that we have all initial parameters for the follow-up study, which will focus on the s-process production in rotating AGB stars. For this study, having a core rotation rate in the AGB phase that is consistent with asteroseismic observations of earlier and later evolutionary phases is crucial.
\end{itemize}

\begin{table*}
\centering
\caption{Summary of all published values for $\nu_{\rm{add}}$ to date}
\begin{tabular}{c|c|c|c}
\hline 
Initial mass (M$_{\odot}$) & $\nu_{\rm{add}}$ (cm$^2$ s$^{-1}$) & Phase & reference \\ 
\hline 
0.84 & 1$\times$10$^3$-1.3$\times$10$^4$ & early red giant & \citet{Eggenberger_2017} \\ 
1.5 & 3 $\times$ 10$^4$ & red giant & \citet{2012eggenberger} \\ 
2.5 & 10$^7$ & core He burning & this work\\
\end{tabular} 
\label{tab:summary}
\end{table*}

\begin{acknowledgements}
We thank the referee, Prof. C. Tout, for his careful reading of the manuscript and comments that greatly improved the clarity and flow of the paper. This work has been supported by the European Research Council (ERC-2012-St Grant 306901, ERC-2016-CO Grant 724560), and the EU COST Action CA16117 (ChETEC). RH acknowledges support from the World Premier International Research Centre Initiative (WPI Initiative), MEXT, Japan. This work is part of the BRIDGCE UK Network (www.bridgce.ac.uk) a UK-wide network established to Bridge Research in the different disciplines related to the galactic chemical evolution and nuclear astrophysics. This paper includes data collected by the Kepler mission. Funding for the Kepler mission is provided by the NASA Science Mission directorate.
\end{acknowledgements}

\bibliographystyle{aa} 
\bibliography{/Users/jdh/Desktop/PhD/Progression_report/references.bib} 

\begin{appendix}
\section{Evolution of rotation from ZAMS to core helium burning}
\label{sec:evo_rot}
\begin{figure}
   \centering
   \includegraphics[width=\linewidth]{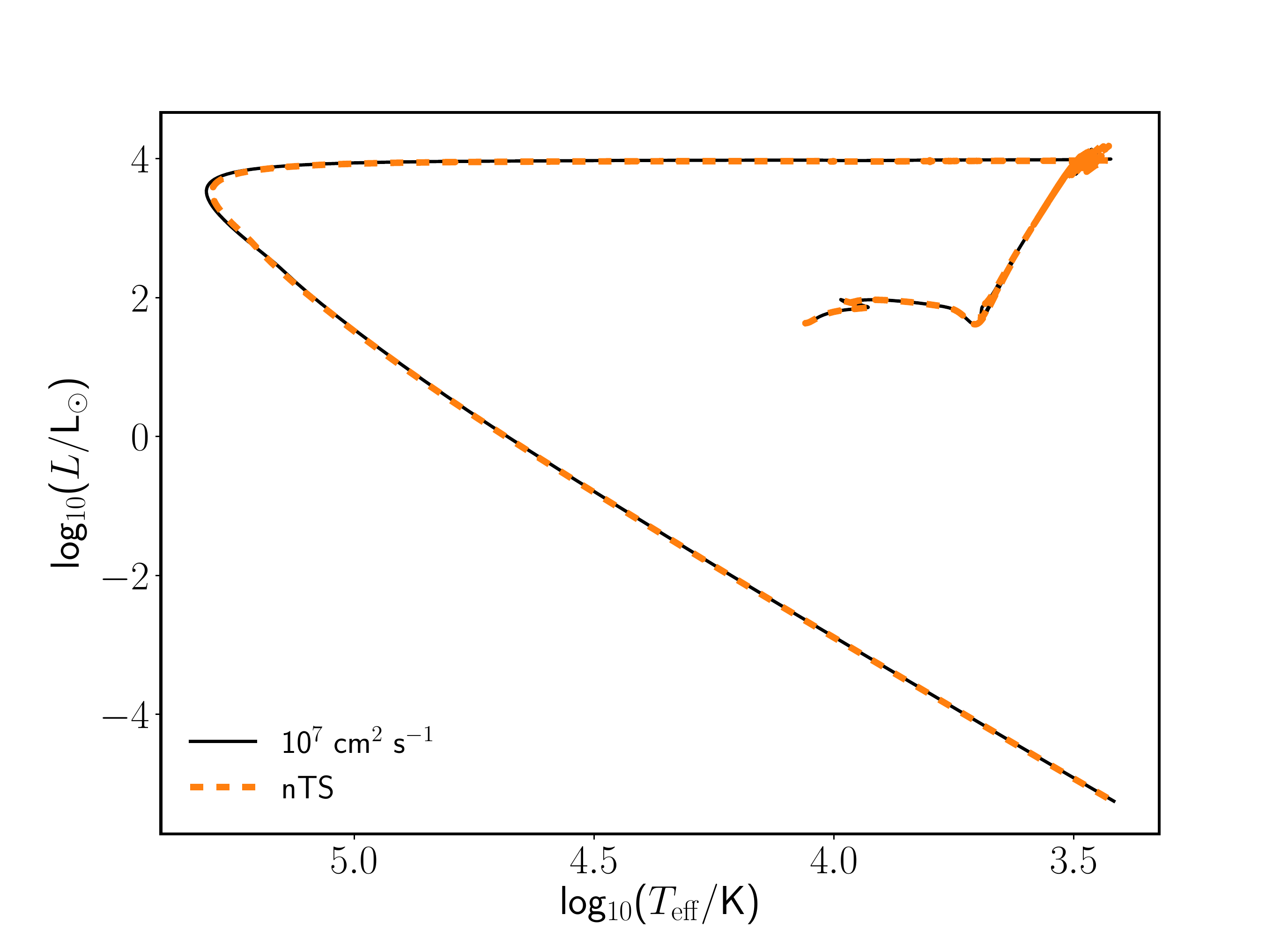} 
   \caption{Hertzsprung-Russell
diagram of 2.5 M$_{\odot}$ models, one without a $\nu_{\rm{add}}$ of 10$^7$  (dashed line) and one with a $\nu_{\rm{add}}$ of 10$^7$ (solid line). Neither model includes the TS dynamo.}
   \label{fig:hrd}
   \end{figure}

\begin{figure}
   \centering
   \includegraphics[width=\linewidth]{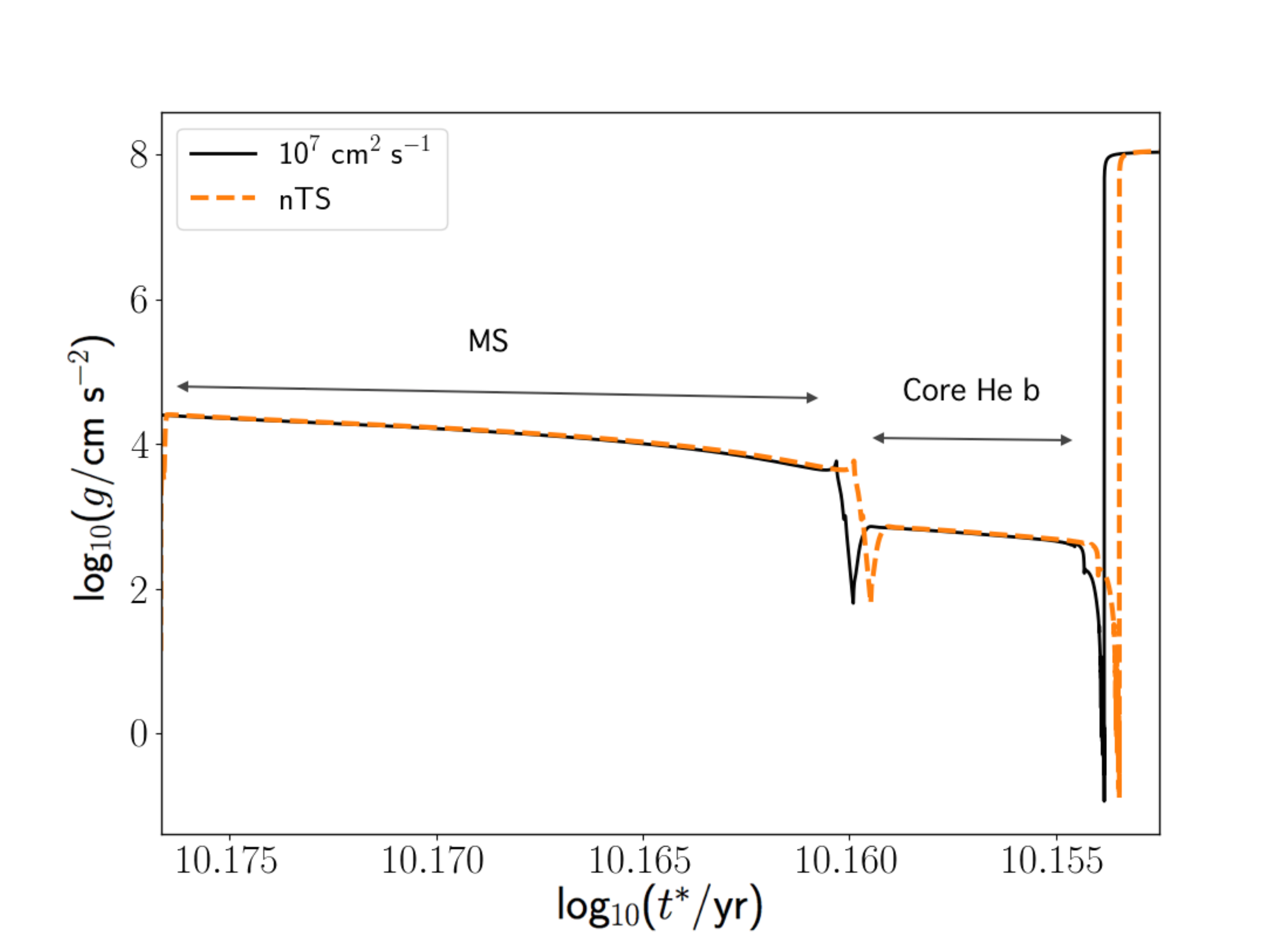} 
   \caption{Time evolution of surface gravity g. Timescale is $t^{*}$=$t_{\rm{WD}}$-$t$), with $t_{\rm{WD}}$ being the age of the star at the end of the calculations. The offset in time comes from a slightly longer white dwarf phase for the 10$^7$ cm$^2$ s$^{-1}$ model compared to the nTS model.}
   \label{fig:log_g_15_101}

   \includegraphics[width=\linewidth]{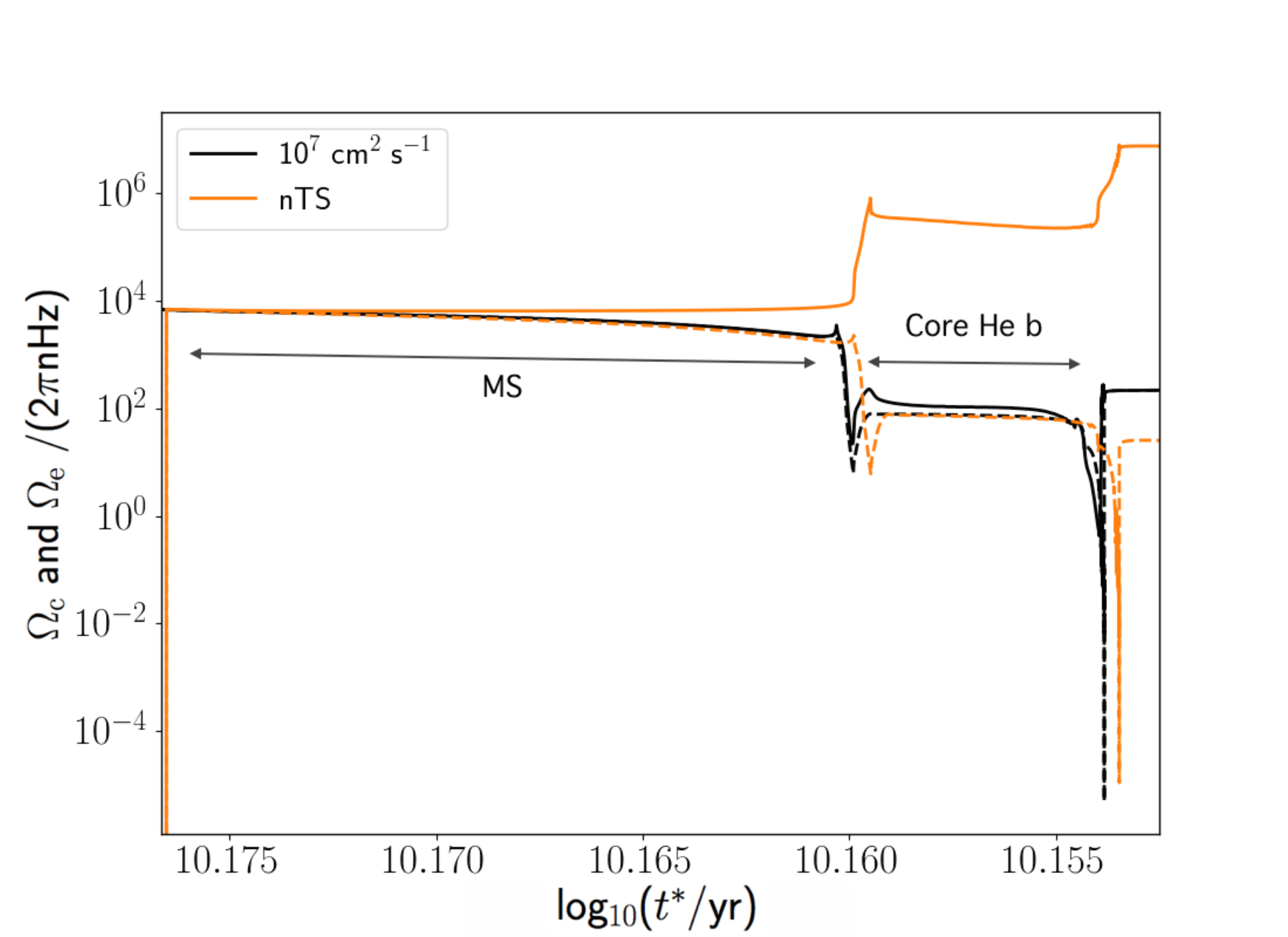} 
   \caption{Coupling made visible: shown here is the evolution of core (solid line) and envelope (dashed) rotation rates from the ZAMS to the start of the AGB phase. Differences between the two models become visible at the start of the hydrogen shell burning phase, where the model without TS dynamo and $\nu_{\rm{add}}$ shows that the core and envelope rotation rates move apart, while the model including a $\nu_{\rm{add}}$ of 10$^7$ cm$^2$ s$^{-1}$ shows the rotation rates are coupled.}
   \label{fig:log_omega_15_101}
   \end{figure}

\begin{figure}
   \centering
   \includegraphics[width=\linewidth]{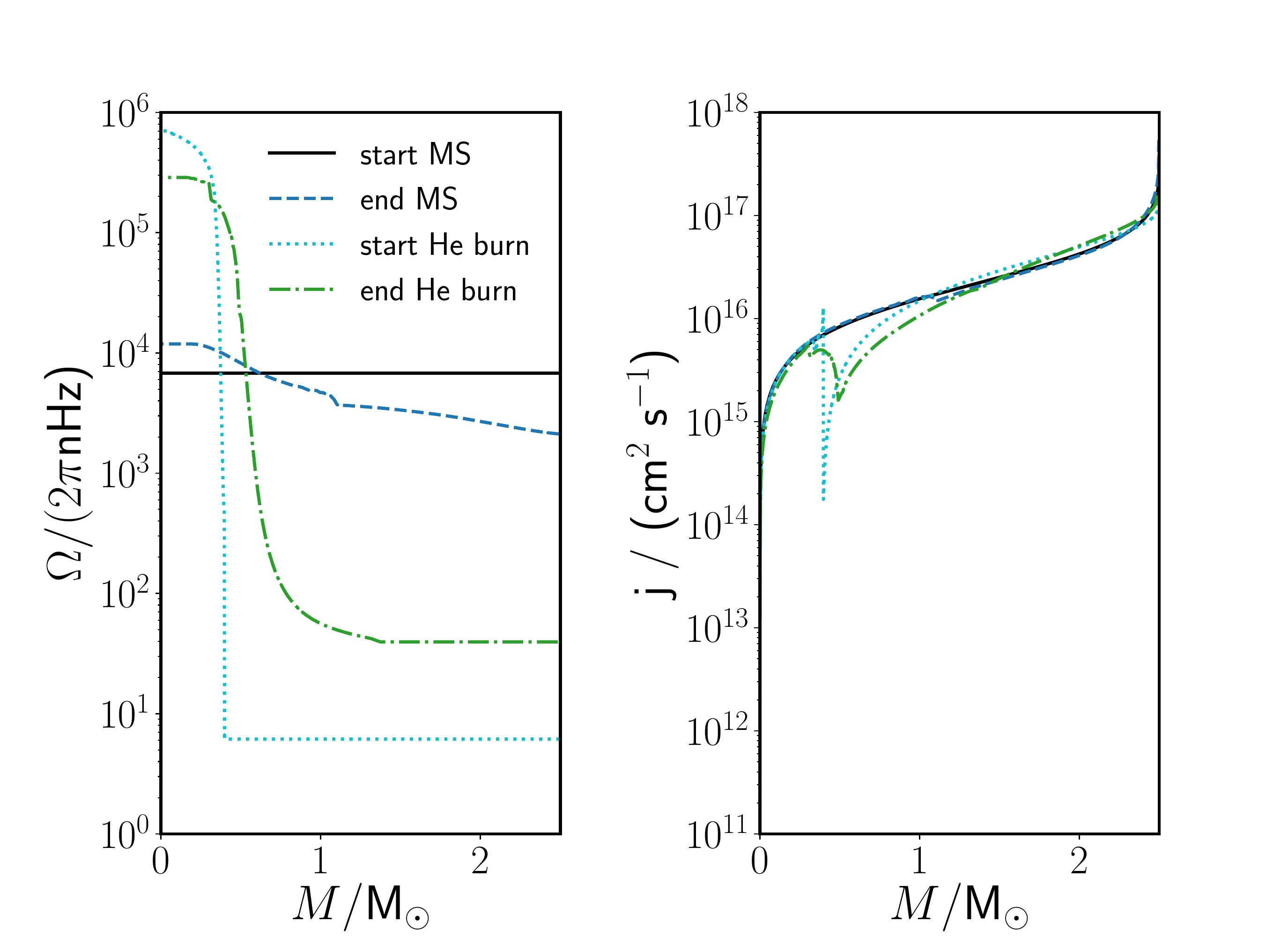} 
   \caption{Angular velocity and angular momentum profiles of the nTS model for four moments as described in the label.}
   \label{fig:omega_101}

   \includegraphics[width=\linewidth]{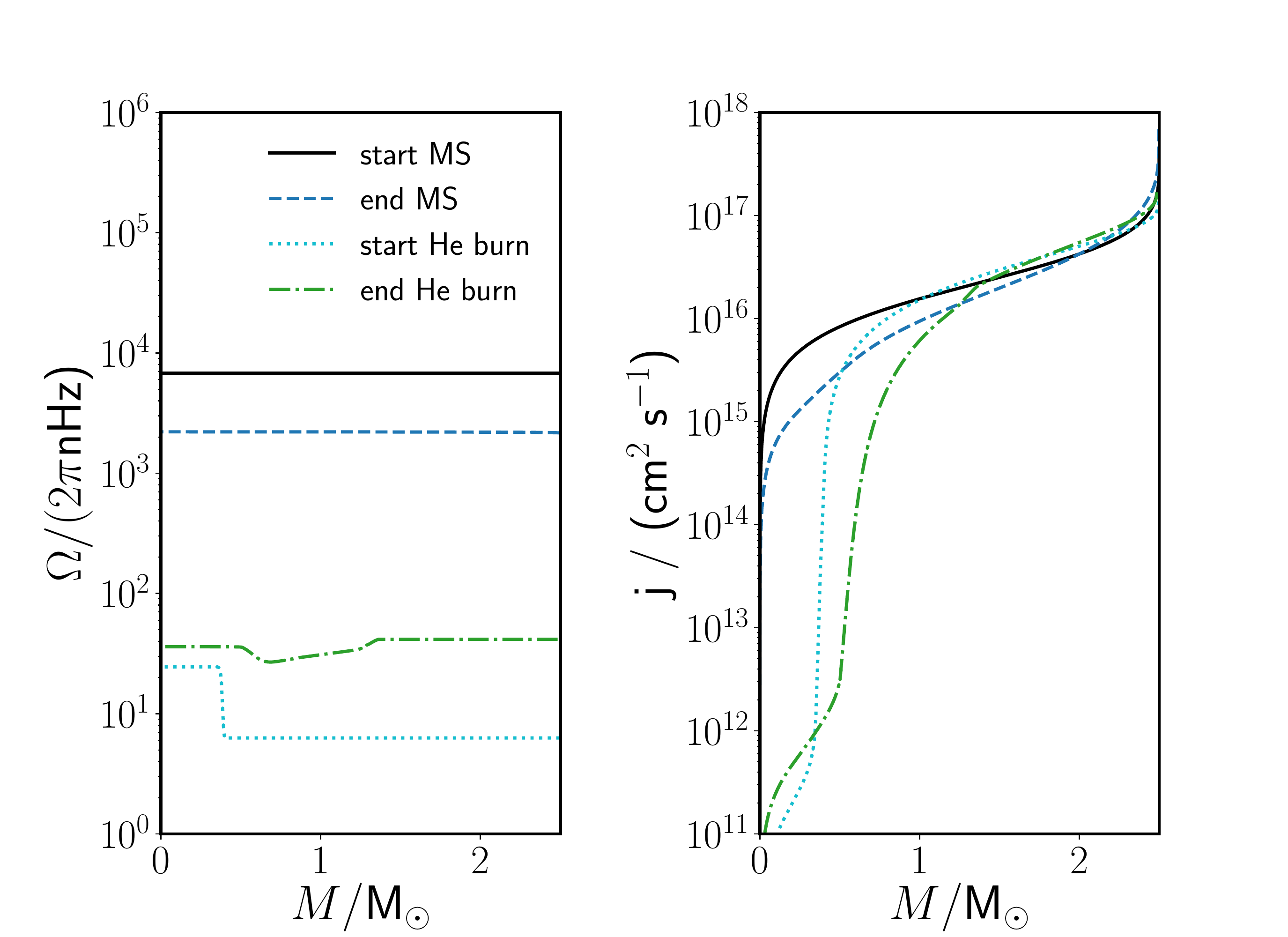} 
   \caption{Same as Fig. \ref{fig:omega_101}, but for the model that includes a $\nu_{\rm{add}}$ of 10$^7$ cm$^2$ s$^{-1}$.}
   \label{fig:omega_15}
   \end{figure}

Before comparing our MESA models to the data points, we introduce the models by discussing their rotational evolution up to the core helium burning phase. The Hertzsprung-Russell diagram (HRD) of two models, one without a $\nu_{\rm{add}}$  (labelled  nTS) and one with a $\nu_{\rm{add}}$ (labelled  10$^7$ cm$^2$ s$^{-1}$) is shown in Fig. \ref{fig:hrd}. The two models do not include the TS dynamo. This figure shows that the two models are comparable. The same is true for the evolution of the surface gravity $g$ shown in Fig. \ref{fig:log_g_15_101}, where log g is shown versus log$_{10}$($t^*$)$\simeq$log($t_{\rm{WD}}$-$t$). In this figure, the horizontal segments of the lines are the core hydrogen (MS) and helium (Core He b) burning phases. The hydrogen shell burning phase takes place in a short amount of time at log$_{10}$($t^*$/yr)$\simeq$10.160, the hydrogen/helium shell burning phase after the core helium burning phase at log$_{10}$($t^*$/yr)$\simeq$10.154. This paper focusses on the core helium burning phase, which starts at log$_{10}$($g$/cm s$^{-2}$)$\simeq$1.8 and a log$_{10}$($t^*$/yr) $\simeq$10.160. Then, in a relatively short amount of time, log$_{10}$($g$/cm s$^{-2})\simeq$2.9 is reached. From there, during the remaining core helium burning phase the log g evolves with a constant slope until log$_{10}$($g$/cm s$^{-2})\simeq$2.4 is reached. This loop is visible in all following log$_{10}$ $g$ vs $\Omega$ figures, where the lower halve of the curves is the long-lasting phase. \\
Figure \ref{fig:log_omega_15_101} shows the time evolution of the angular velocity of core $\Omega_{\rm{c}}$ (solid lines) and envelope $\Omega_{\rm{e}}$ (dashed lines) from the start of the main sequence to the start of the AGB phase. During the core burning phases, the rotation rates of core and envelope are close to constant in both models, with the model including $\nu_{\rm{add}}$ showing a near solid body rotation trend during the main sequence. The nTS model, however, shows large differences between core and envelope rotation rates during the shell burning phases. These phases are characterised with core contraction and envelope expansion \citep[also known as the mirror principle\footnote{This principle is not a physical law, but an empirical observation confirmed by numerical simulations. It states that when a region within a burning shell contracts the region outside the shell will expand, and vice versa.}, see][]{book_kipp}, resulting in a steeply increasing core rotation rate and steeply decreasing envelope rotation rate. \\
The model including $\nu_{\rm{add}}$ shows different trends during the shell burning phases. The coupling provided by $\nu_{\rm{add}}$ allows for transport of angular momentum even when the core is contracting. As a result, the core rotation rate follows the trends of the envelope rotation rate and decreases during the shell burning phases (orange lines in Fig. \ref{fig:log_omega_15_101}). This trend is as observed by \citet{Aerts2017}, who compare a compilation of rotation rates of main sequence stars to the rotation rates of more evolved stars by \citet{2012Mosser}. They find that there must be a drop in core rotation before or during the end of hydrogen and the start of helium core burning phases.\\
The details of the angular velocity $\Omega$ and corresponding angular momentum $j$ profiles from core to surface are given in Figs. \ref{fig:omega_101} and \ref{fig:omega_15}. Both figures show this profile at four moments in the evolution: the start and end of the main sequence and the start and end of the core helium burning phase. The solid body start of the models is visible in both figures, and from there the differences appear. As mentioned before, the angular velocity of the core and envelope in the model without $\nu_{\rm{add}}$ (left panel of Fig. \ref{fig:omega_101}) evolve separately and oppositely due to the mirror principle. This effect is already visible at the end of the main sequence, and results in a difference between core and envelope rotation rate of several orders of magnitude at the end of the core helium burning phase. In the right panel the $j$ profiles are shown. A decrease in $j$ in a region during a certain phase indicates transport of angular momentum. A sharp feature is usually the outer edge of a convective zone, which creates a barrier for transport of angular momentum. The general lack of transport of angular momentum in the nTS model is visible in the $j$ profiles of Fig. \ref{fig:omega_101}, because they largely overlap. \\ 
When an additional viscosity of $\nu_{\rm{add}}$=10$^7$ cm$^2$ s$^{-1}$ is added, the differences between core and envelope angular velocity are smaller than in the nTS model (left panel of Fig. \ref{fig:omega_15}). The whole star is close to solid body rotation up to the end of the core helium burning phase, as also shown in Fig. \ref{fig:log_omega_15_101}. In this model a large amount of angular momentum is transported out of the core between the end of the main sequence and the start of the core helium burning phase (right panel of Fig. \ref{fig:omega_15}).  This efficient transport is also able to overcome the edge of convective regions, resulting in a lack of sharp features in the $j$-profiles. The transport continues during the core helium burning phase, creating a short moment at the end of the core helium burning phase when the convective envelope rotates at a higher angular velocity than the rest of the star. \\

\section{Model uncertainties}
\label{sec:model_unc}
\begin{figure}
   \centering
   \includegraphics[width=\linewidth]{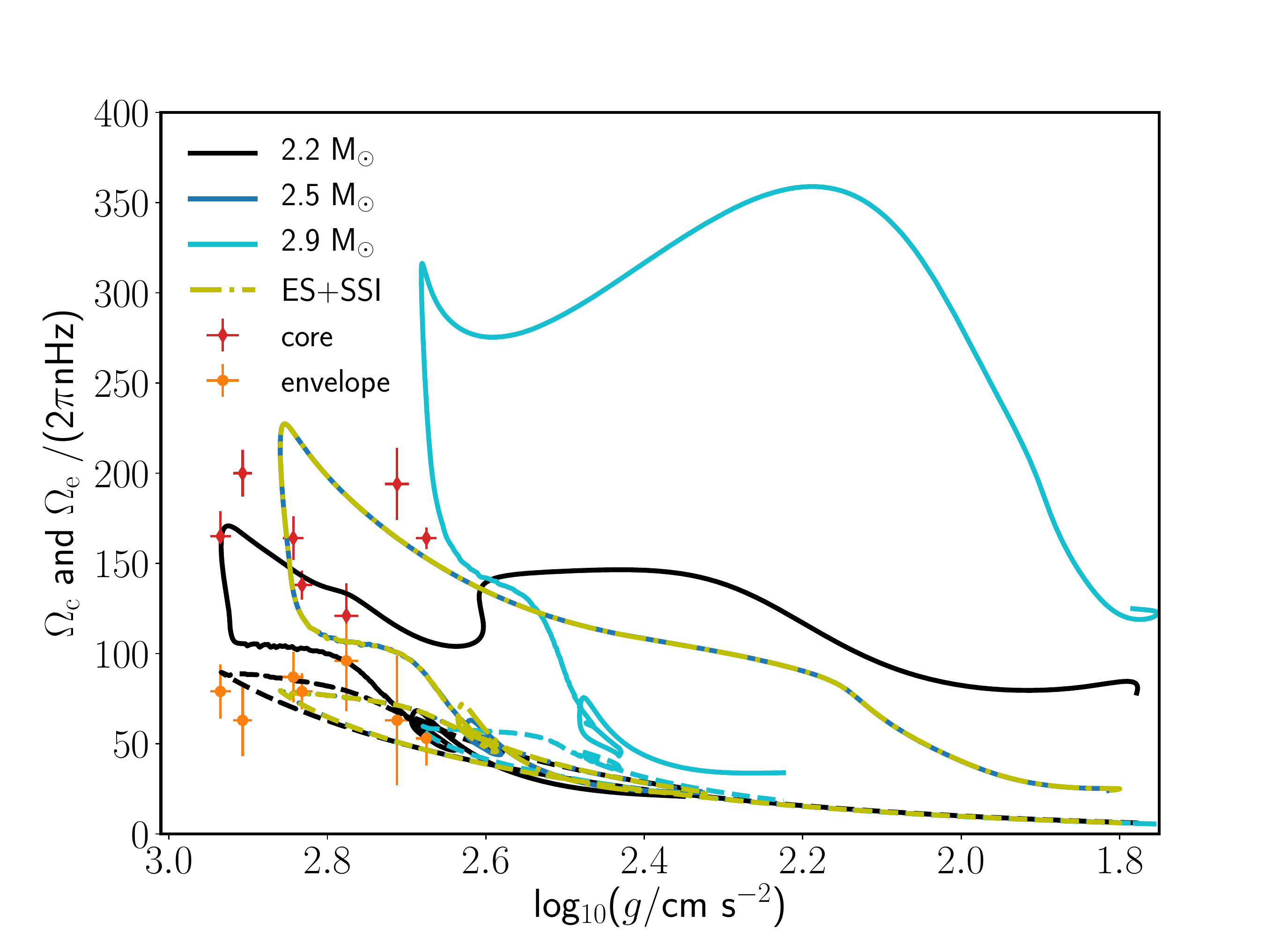}
   \caption{Model uncertainties: the first three models presented here have been calculated with the best fit parameters (50 km s$^{-1}$,  10$^7$ cm$^2$  s$^{-1}$), while the initial mass is varied. The fourth model includes only the ES and SSI instabitity. The models labelled  `2.5 M$_{\odot}$' and `ES+SSI' overlap.}
   \label{fig:omega_artvisc_3masses}%
   \end{figure}

In Sect. \ref{sec:artvisc} we were not able to mass the data points at the highest surface gravities corresponding to KIC5184199 and KIC4659821. Here, we show that this is a consequence of setting the initial mass to 2.5 M$_{\odot}$. When matching the initial mass to the masses listed in Table \ref{tab:deheuvels}, we can indeed match the highest surface gravities, as shown in Fig. \ref{fig:omega_artvisc_3masses}. For all models in this comparison, we use $\nu_{\rm{add}}$ = 10$^7$ cm$^2$  s$^{-1}$. The model with the lowest initial mass (2.2 M$_{\odot}$) reaches the higher surface gravities of the two data points earlier unreached. These two data points correspond to the observations of stars with initial masses of 2.18 $ \pm$ 0.23 and 2.21 $\pm$ 0.18 M$_{\odot}$, indeed matching the lower initial mass of 2.2 M$_{\odot}$. When comparing the model with the highest initial mass (2.9 M$_{\odot}$) to the data points, we find that the star with the highest mass, KIC7581399, of 2.90 $\pm$ 0.34 M$_{\odot}$, has a log$_{10}$($g/$cm s$^{-2}$) = 2.843 $\pm$ 0.013 and is  located on the 2.5 M$_{\odot}$ model. This might imply that the actual mass of KIC7581399 is located near the lower end of the error margin.\\
The implementation of rotation in MESA allows for the inclusion and exclusion of individual rotationally induced instabilities. The dynamical instabilities (DSI and SH) are not part of the GENEC models as published by Eggenberger et al. (2012, 2017). Here we investigate their effects on the transport of angular momentum in the MESA models studied in this paper. To test this, we calculated an extra model with an initial mass of 2.5 M$_{\odot}$ and $\nu_{\rm{add}}$ = 10$^7$ cm$^2$  s$^{-1}$ with only the ES and SSI included, and added this model to Fig. \ref{fig:omega_artvisc} with the label `ES+SSI'. The overlap of this model and the 2.5 M$_{\odot}$ model, which also  includes  the dynamical instabilities, shows that the SH and DSI do not contribute to the transport of angular momentum. \citet{P.V.F.Edelmann2017} have already shown issues with the 1D implementation of the DSI in stellar evolutionary codes, and therefore being able to exclude this instability in studies on angular momentum transport reduces the uncertainties of our results. They also confirm that the settings of the GENEC models are satisfactory.\\

\begin{figure}
   \centering
   \includegraphics[width=\linewidth]{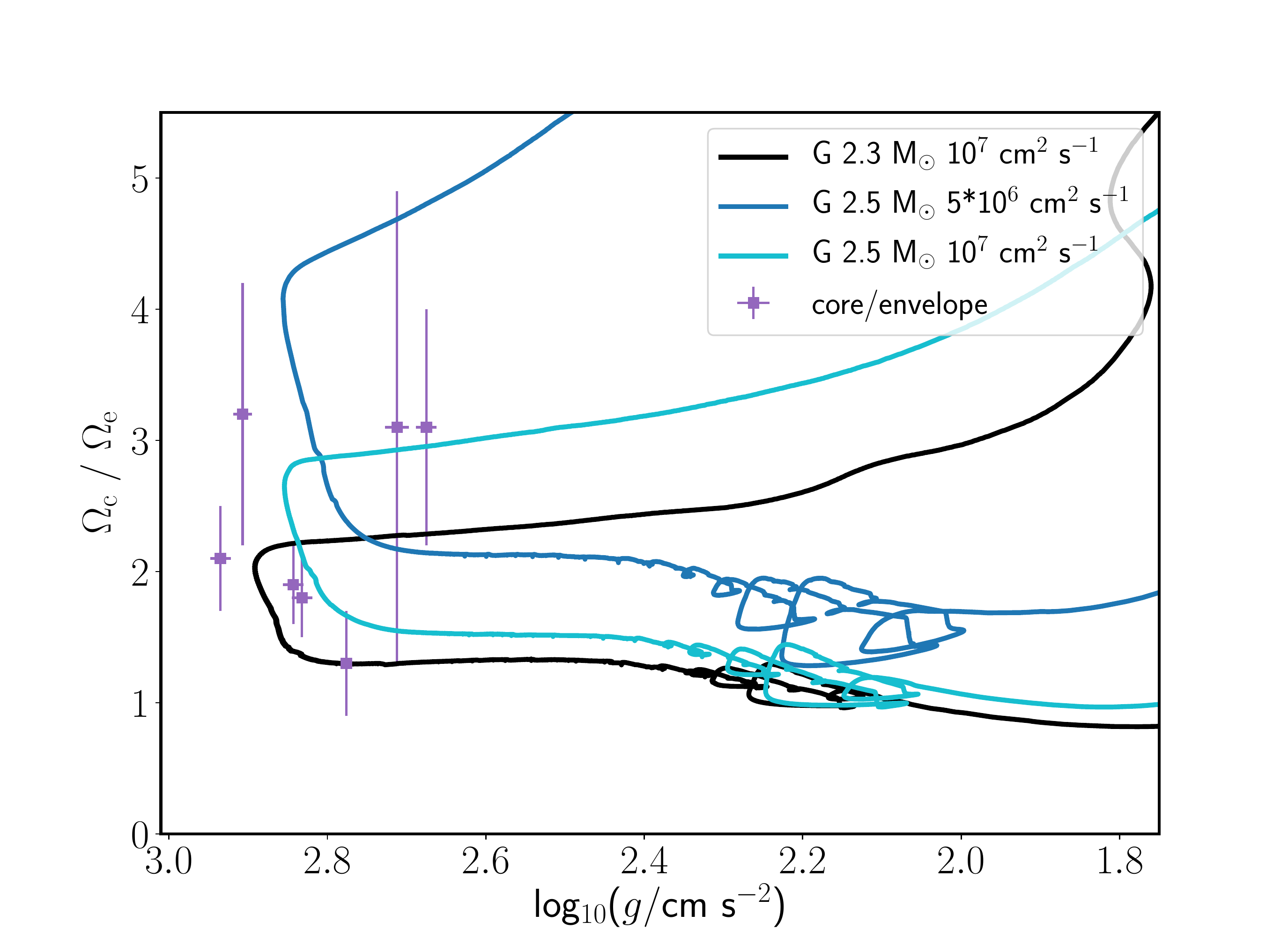}
   \caption{Code comparison: the models presented here are calculated with GENEC to show our conclusions are independent from evolutionary code.}
   \label{fig:ratio_MandG}%
   \end{figure}

In Sect. \ref{sec:introduction} we mention that we ran the best fit models of Eggenberger et al. (2012, 2017), and found  $\nu_{\rm{add}}$ values that were similar to theirs  to explain the observations. Here, we show GENEC models \citep[see][for a description of this code and their implementation of rotation]{2008_geneva_eggenberger} calculated to match the MESA models of this study. Three GENEC models are shown in Fig. \ref{fig:ratio_MandG}, with  their  $\nu_{\rm{add}}$ and initial mass, as labelled. The same trends can be identified in these models as in the MESA models of earlier sections: when the initial mass is reduced, the data points at high surface gravities can be reached. A $\nu_{\rm{add}}$ of 10$^7$ cm$^2$ s$^{-1}$ provides a better fit than 5$\times$10$^6$ cm$^2$ s$^{-1}$. Therefore, our conclusions are code independent.

\section{Rotation near the core}
\label{sec:nearcore}
\begin{figure}
   \centering
   \includegraphics[width=\linewidth]{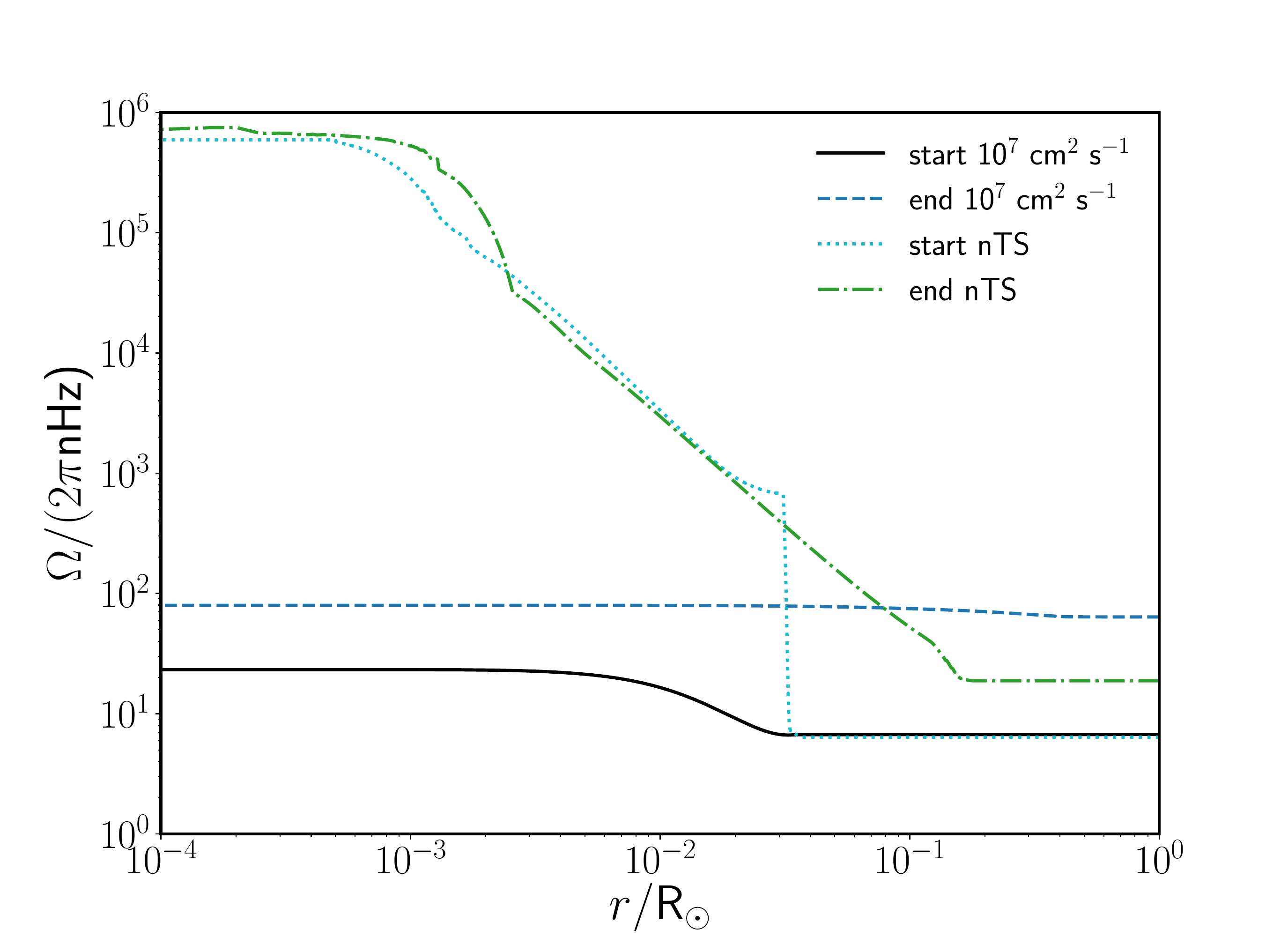}
   \caption{For comparison with Fig. 5 of \citet{Deheuvels2015}. The region of interest is  between $r$/R$_{\odot}$ of 10$^{-3}$ and 10$^{-2}$.}
   \label{fig:omega_norm_radius}%
   \end{figure}

As mentioned in Sect. \ref{sec:7KIC}, the numbers in the  core rotation rates column in Table \ref{tab:deheuvels} are actually `near core' rotation rates. Their location is 0.1--1\% of the normalised radius away from the most central point, see Fig. 5 in \citet{Deheuvels2015}. In this region the obtained rotation rate is constant despite the noise in this figure. In Fig. \ref{fig:omega_norm_radius} we show a similar figure for the nTS and 10$^7$ cm$^2$ s$^{-1}$ models, where the rotation rate at the start and end of the core helium burning phase is shown. We see that the model including the extra $\nu_{\rm{add}}$ shows a constant trend in the region of interest at both times, as needed for the comparison to the data of \citet{Deheuvels2015}. However, the nTS model shows a strong decrease in this region, providing another argument against these standard models.
\end{appendix}

\end{document}